\documentclass[submission]{eptcs}
\providecommand{\event}{TTC 2011} % Name of the event you are submitting to

\usepackage[english]{babel}
\addto\extrasenglish{%
	
}

\usepackage{csquotes}
\usepackage{graphicx}
\usepackage{hyperref}
\usepackage{listings}
\usepackage{subfig}
\usepackage{xspace}

\date{\year=2011\month=3\day=27\today}

\newcommand{\node}[1]{{\sf #1}}
\newcommand{\Firm}{\textrm{\textmd{\textsc{Firm}}}\xspace}

\title{Compiler Optimization: A Case for the Transformation Tool Contest}
\author{Sebastian Buchwald \quad \quad Edgar Jakumeit
\institute{Karlsruhe Institute of Technology (KIT)}
\email{buchwald@kit.edu \quad \quad \phantom{~~~~~~~~~~~~~~~~}}
}
\def\titlerunning{Compiler Optimization Case}
\def\authorrunning{Sebastian Buchwald and Edgar Jakumeit}
\begin{document}
\maketitle

%\begin{abstract}
%This is a sentence in the abstract.
%This is another sentence in the abstract.
%This is yet another sentence in the abstract.
%This is the final sentence in the abstract.
%\end{abstract}

\section{Introduction}
An optimizing compiler consists of a front end parsing a textual programming language into an intermediate representation (IR), a middle end performing optimizations on the IR, and a back end lowering the IR to a target representation (TR) built of operations supported by the target hardware.
In modern compiler construction graph-based IRs are employed.
Optimization and lowering tasks can then be implemented with graph transformation rules.

\vspace{2mm}
\noindent The participating tools are required to solve the following challenges:
\begin{description}
	\item[Local optimizations] replace an IR pattern of limited size and only local graph context by another semantically equivalent IR pattern which is cheaper to execute or which enables further optimizations.
		For instance, given the associativity of the '+'-operator, the transformation $(x+1)+2 \Rightarrow x+(1+2) \Rightarrow x+3$ can be performed by local optimizations.
	\item[Instruction selection] transforms the intermediate representation into the target representation.
		While both representations are similar in structure, the TR operations are often equivalent to a small pattern of IR operations, i.e. one TR operation covers an IR pattern (it may be equivalent to several different IR patterns).
		Thus instruction selection can be approached by applying rules rewriting IR patterns to TR nodes,
		until the whole IR graph was covered and transformed to a least cost TR graph.
\end{description}

\noindent The primary purpose of these challenges is to evaluate the participating tools regarding \emph{performance}.
The ability to apply rules in parallel should be beneficial for the instruction selection task due to the locality of the rules and a high degree of independence in between the rules.
It could be advantageous for the optimization task, too, but intelligent traversal strategies might have a higher impact there.
Secondary aims are fostering tool interoperability regarding the GXL standard format and testing the ability to visualize medium sized real world graphs.

We provide the following resources on the case website~\cite{compilerweb}:
\begin{itemize}
	\item This case description.
	\item Input graphs of various sizes in GXL format.
	\item The needed meta models are contained in the GXL files.
\end{itemize}

\section{A Graph-Based Intermediate Representation}

The intermediate representation to be used is a simplified version of the graph-based intermediate representation \Firm\footnote{\url{www.libfirm.org}}.
Within \Firm, nodes represent basic operations, e.g.\ loading a value from an address or adding two values.
Edges indicate dependencies between nodes, e.g.\ an \node{Add} depends on its two operands.
Each operation is located within a \node{Block}\footnote{also known as basic block}.
For the represented program all nodes within the same \node{Block} are executed together, i.e.\ if one node within a \node{Block} is executed then all nodes within the \node{Block} must be executed.

\begin{figure}[htbp]
	\begin{minipage}[c]{0.4\textwidth}
		\centering
		\includegraphics[width=1.7in]{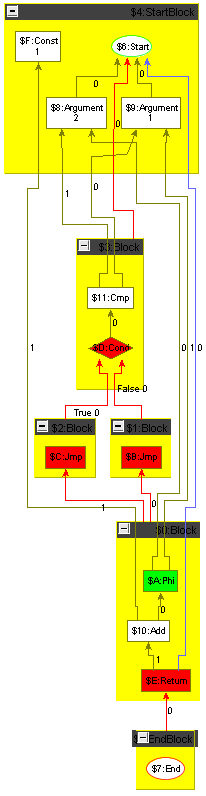}
	\end{minipage}
	\begin{minipage}[c]{0.6\textwidth}
		\centering
		\includegraphics[width=2.7in]{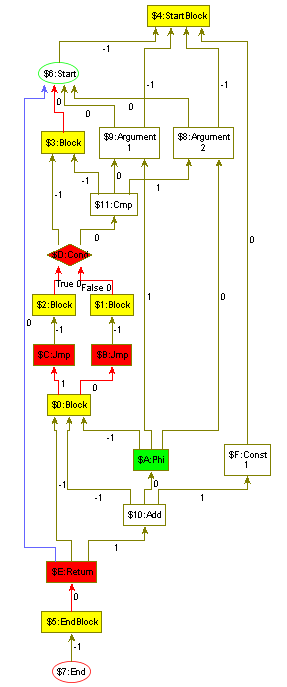}
	\end{minipage}
	\caption{Program graph of a minimum plus one function, with block containment visualized as containment left, and  the plain graph right.}
	\label{fig:max}
\end{figure}

\noindent \autoref{fig:max} shows the program graph of our reference solution~\cite{compileroptimizationsolutiongrgennet} for the following C-function:
\begin{lstlisting}
int MinPlus(int x, int y)
{
	int min;

	if (x < y) {
		min = x;
	}
	else {
		min = y;
	}

	return min + 1;
}
\end{lstlisting}
The program execution starts at \node{Start}, which also produces the initial memory state and the given \node{Argument}s.
\node{Start} and the \node{Argument}s belong to the special \node{StartBlock}.
The \node{Cmp} compares the arguments and \node{Cond} represents a conditional jump depending on the result of the comparison.
After executing the \node{Jmp} of the then- or else-\node{Block} the program execution continues at \node{Block} \$0.
The \node{Phi} originates form the static single assignment form~\cite{ssa} which is required for a concise graph-based IR.
It chooses one of its operands depending on the previously executed block.
For instance, the \node{Phi} selects \node{Argument} $2$ if the \node{Cond} was evaluated to \node{False} and \node{Argument} $1$ if the \node{Cond} was evaluated to \node{True}.
The \node{Return} returns the value selected by the \node{Phi}.
The end of the program execution is represented by the special \node{EndBlock} which by convention contains exactly one \node{End}.

Each node has ordered outgoing edges, which is indicated by the \node{position} attribute of the edges.
Edges representing the containment to a Block have \node{position} $-1$; the operands start at position $0$.
There are several edge types:
\begin{description}
	\item[Dataflow] Models the flow of data from an operation to another one.
	\item[Memory] Memory edges are used to ensure an order for memory operations.
	\item[Controlflow] Models possible execution paths through the program.
	\item[True] Control flow if a condition jump is evaluated to true.
	\item[False] Control flow if a condition jump is evaluated to false.
	\item[Keep] Needed to model infinite loops, see description of \node{End}.
\end{description}
The direction of the edges is reversed to the direction of the flow (it follows the dependencies).
For instance, \node{start} has only incoming edges.

The following list describes all node types (the $^\ast$ marks the types which are of relevance for the instruction selection task):
\begin{description}
	\item[Block] A basic block.
		All outgoing edges point to possible execution predecessors.
		The incoming Dataflow edges define the contained operations.
		In the figures the contained operations are shown as containment in the Block node for easier readability, the right subfigure of \autoref{fig:max} displays the plain graph structure underneath.
	\item[StartBlock] The start block.
	\item[Start] The starting point of the program.
		Produces initial control flow and memory state.
	\item[Argument] The arguments of the program.
		The \node{position} attribute indicates which argument is represented.
	\item[EndBlock] The end block.
	\item[End] Nodes which are not reachable by the \node{End} can be removed from the graph.
		Since infinite loops may never reach the \node{EndBlock}, \node{Keep} edges are inserted to prevent the removal of such a loop.
	\item[Phi] A \node{Phi} selects the operand of the previously executed \node{Block} from the operands available.
		This implies that the number of operands of a \node{Phi} node must be equivalent to the number of control flow predecessors of the \node{Block} it is contained in.
		Furthermore the \node{position} attributes of the operand edges of the \node{Phi} node must be equal to the \node{position} attributes of the corresponding control flow edges of the containing \node{Block} node.
	\item[Jmp$^\ast$] A jump to a \node{Block}.
	\item[Cond$^\ast$] A conditional jump.
		The operand must be a \node{Cmp} or another node producing a value $\in\{0,1\}$, e.g. a \node{Const}.
	\item[Return] Returns a value.
		The first operand must be the current memory state and the second operand the return value.
	\item[Const$^\ast$] A constant value.
		The value of the constant of integer type is presented by the \node{value} attribute.
		A \node{Const} is always located in the \node{StartBlock}.
	\item[SymConst$^\ast$] A symbolic constant.
		A \node{SymConst} is used to represent the address of some global data, e.g.\ arrays.
		The \node{symbol} attribute of string type represents the name of the global data.
		A \node{SymConst} is always located in the \node{StartBlock}.
	\item[Load$^\ast$] Load a value from a given address.
		The first operand is the current memory state and the second operand is the address.
		The \node{Load} produces a new memory state.
		The \node{volatile} attribute indicates whether the value at the corresponding memory address can be changed by some other thread.
		For instance, two consecutive \node{volatile} \node{Load}s from the same address cannot be merged.
	\item[Store$^\ast$] Stores a value to a given address.
		The first operand is the current memory state, the second operand is the address and the third operand the value that should be stored.
		Similar to the \node{Load}, a \node{Store} also has a \node{volatile} attribute.
	\item[Sync] Synchronize multiple memory operations.
		The \node{Sync} is used to represent that some memory operations are not in any particular order.
		For instance, \node{Load}s of different array elements $a[i]$, $a[i+1]$ may not be ordered.
	\item[Not$^\ast$] Bitwise complement.
	\item[Binary$^\ast$] An operation with two operands.
		Each binary operation has two boolean attributes: \node{associative} and \node{commutative}.
		The following binary operations are supported:
		\begin{itemize}
			\item \node{Add$^\ast$}
			\item \node{Sub$^\ast$}
			\item \node{Mul$^\ast$}
			\item \node{Div$^\ast$}
			\item \node{Mod$^\ast$}
			\item \node{Shl$^\ast$}---shift left, fill up with zero
			\item \node{Shr$^\ast$}---shift right, fill up with zero
			\item \node{Shrs$^\ast$}---shift right signed, fill up with sign bit ($1$ if negative, $0$ otherwise)
			\item (bitwise) \node{And$^\ast$}
			\item (bitwise) \node{Or$^\ast$}
			\item (bitwise) \node{Eor$^\ast$}---exclusive or
			\item \node{Cmp$^\ast$}---compare two values.\\
				The \node{relation} attribute indicates the checked relation, i.e.\ one of FALSE, GREATER, EQUAL, GREATER\_EQUAL, LESS, NOT\_EQUAL, LESS\_EQUAL, and TRUE.
		\end{itemize}
\end{description}

% TODO Upload the test cases and provide a link

\section{Getting started: Verifier}
To get started you can create some rules that check your graph for validity.
This may include the following checks:
\begin{itemize}
	\item There is only one \node{Start}.
	\item There is only one \node{End}.
	\item A \node{Dataflow} edge to a \node{Block} always has \node{position} $-1$.
	\item Constants are only located in the \node{StartBlock}.
	\item A \node{Phi} has as many operand \node{Dataflow} edges as the \node{Block} it is located in \node{Controlflow} predecessors.
	\item The operand \node{Dataflow} edges of a \node{Phi} are linked to the \node{Controlflow} edges of the \node{Block} it is contained in via the \node{position} attribute in those edges - are they ascending from 0 on without gaps?
\end{itemize}
The verifier is not part of the challenge,
but it might be helpful in checking the correctness of your solution of the following tasks.

\section{Task 1: Local Optimizations}
The first task is to optimize programs using local optimizations, i.e.\ rules with a pattern of fixed size and only local graph context.

\subsection{Constant Folding}
Constant folding means to evaluate an operation with only constants operands, e.g.\ to transform $1+2$ into $3$.
Constant folding of data flow operations is a straight-forward task, a good deal more complicated is constant folding including control flow, replacing conditional jumps with constant operands by unconditional jumps.

\begin{figure}[htbp]
	\centering
	\subfloat[After folding the Cmp.]{\label{fig:cmpfold}
		\includegraphics[width=2.0in]{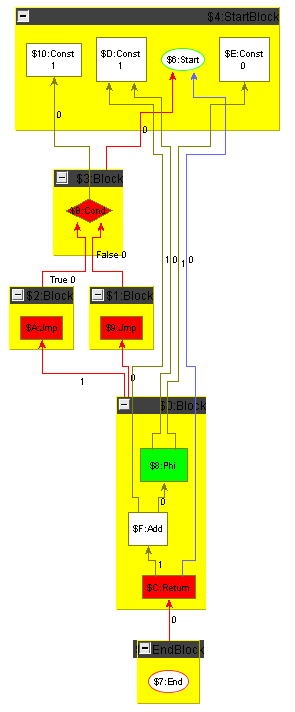}
	}
	\qquad
	\subfloat[After folding the Cond.]{\label{fig:condfold}
		\includegraphics[width=2.0in]{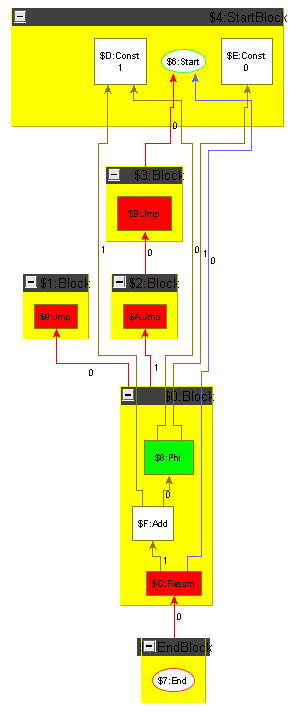}
	}
	\caption{Program graph of the minimum plus one function.}
	\label{fig:cmpcondfold}
\end{figure}

\begin{figure}[htbp]
	\centering
	\subfloat[After elimination of the unreachable block and the Phi operand.]{\label{fig:unreachablefold}
		\includegraphics[width=2.0in]{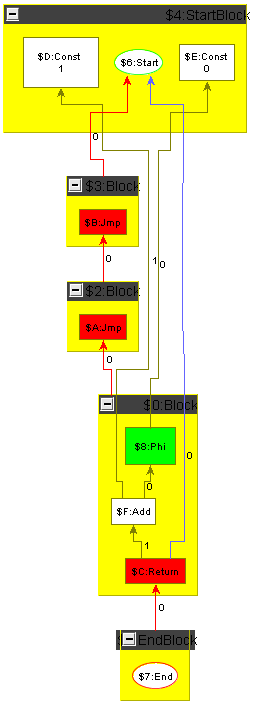}
	}
	\qquad
	\subfloat[After empty block removal.]{\label{fig:emptyfold}
		\includegraphics[width=2.0in]{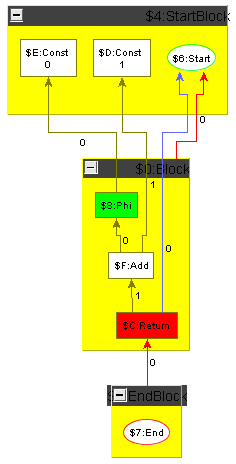}
	}
	\caption{Program graph of the minimum plus one function.}
	\label{fig:removals}
\end{figure}

\begin{figure}[htbp]
	\centering
	\subfloat[After folding the Phi.]{\label{fig:phifold}
		\includegraphics[width=2.0in]{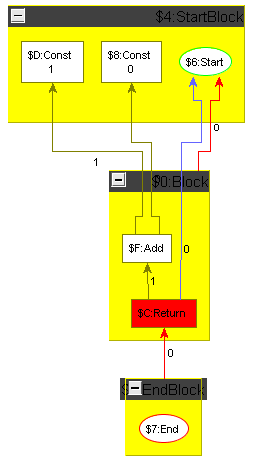}
	}
	\qquad
	\subfloat[After folding the Add.]{\label{fig:addfold}
		\includegraphics[width=1.7in]{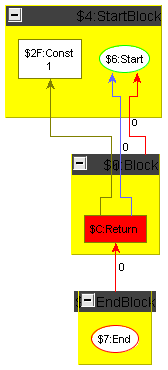}
	}
	\caption{Program graph of the minimum plus one function.}
	\label{fig:allfolded}
\end{figure}

To ease the challenge for participants not knowledgeable in compiler construction we show how to carry out constant folding including control flow on an example graph.
The example graph is the minimum plus one graph introduced in the previous section, but with the arguments replaced by constants; it is shown in the following section, \autoref{fig:target}, left side.
Folding starts with the \node{Cmp} node comparing two constant values: it is folded to a constant giving the result of the comparison, which is $1$ denoting $true$ in our case; the result is displayed in \autoref{fig:cmpfold}.
Then the \node{Cond} node depending only on the constant created in the previous step is folded, i.e. replaced by an unconditional \node{Jmp}; the result is shown in \autoref{fig:condfold}.
In the next step, we remove the unreachable false \node{Block}.
When a block is removed, the \node{Phi}s in the blocks succeeding it must get adapted, i.e.\ the operands which resulted from executing that block must get removed.
Additionally the \node{position} attributes of the control flow edges and of the data flow edges of the Phi operands must be decremented from the correct position on.
The result is shown in \autoref{fig:unreachablefold}.
Now the empty blocks denoting a useless jump cascade can get removed and the \node{Phi} folded;
a \node{Phi} with only one dependency edge is superfluous as there is no decision to be taken at runtime any more, it can get replaced by relinking its users directly to its input value.
Removing the empty blocks first and folding the Phi afterwards (we could reverse this or do it in parallel),
we reach via \autoref{fig:emptyfold} the situation given in \autoref{fig:phifold}.
Folding the \node{Add} node now only depending on two constants, we reach the end result of the optimization, a function returning the constant value $1$.

% TODO
To make the tools comparable, there will be a test suite of multiple programs (in GXL format) that can be optimized by local optimizations.

\subsection{Extension}
You may challenge the other participants by adding your own test programs to the tests suite.
The only requirement is that the corresponding optimization is a local optimization.
This offers the opportunity to highlight strong points of your tool.

\section{Task 2: Instruction Selection}

The second task is to perform instruction selection, i.e.\ to transform the intermediate representation into a target-specific representation.
We assume a very simple target architecture resulting in a TR which is structurally nearly identical to the IR, with the sole exception of \enquote{immediates} which may be encoded directly in the target instructions\footnote{More precisely, we assume some kind of simple RISC machine; the more interesting case of a CISC machine, where target operations can contain a memory access, giving rise to non-trivial IR patterns to be covered, is too complicated for this contest.}

All operations that must be transformed by instruction selection are marked with a '$^\ast$'.
For each operation \node{Op} there is a target-specific operation \node{TargetOp} representing an operation on registers, e.g.\ \lstinline{R1 = add R2, R3}.
For each binary operation \node{Op} there is an additional target-specific operation \node{TargetOpI}, which has one constant operand, e.g.\ \lstinline{R1 = add R2, 42}.
It represents an IR pattern \node{Op(x, Const)}.
The value of the constant is called \enquote{immediate} and is stored in the additional \node{value} attribute.
For non-commutative binary operations the \node{Const} must be at the outgoing edge with \node{position} $1$.
For \node{Load} and \node{Store} there are in addition target-specific operations \node{LoadI} and \node{StoreI} with an additional attribute \node{symbol} which holds a symbolic constant; they represent IR patterns \node{Load(mem, SymConst)} and \node{Store(mem, SymConst, x)}.

\begin{figure}[htbp]
	\begin{minipage}[c]{0.5\textwidth}
		\centering
		\includegraphics[width=2.7in]{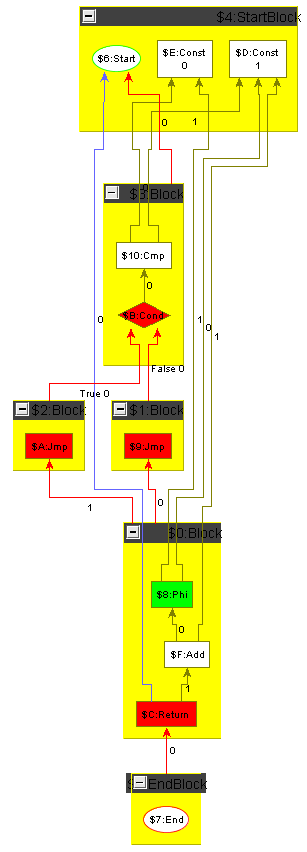}
	\end{minipage}
	\begin{minipage}[c]{0.5\textwidth}
		\centering
		\includegraphics[width=2.3in]{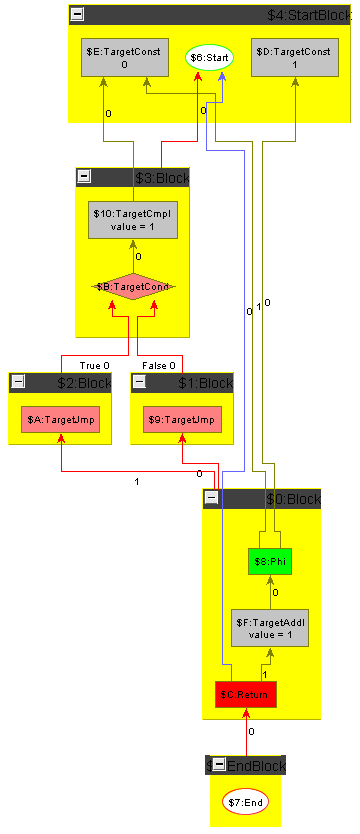}
	\end{minipage}
	\caption{Program graph before and after instruction selection.}
	\label{fig:target}
\end{figure}

The goal of this task is to perform optimal instruction selection with respect to the number of resulting operations.
This means an operation with immediate is always better than the same operation without immediate.
\autoref{fig:target} shows the minimum plus one function on constant values not optimized away by constant folding, before and after instruction selection.
Performance hint: the setup given allows to process all operations with immediate and then all operations without immediate in parallel.

\section{Evaluation criteria}
For each task tackled the authors should give a short overview of their solution.
Furthermore, the solution should contain some statements regarding the following criteria:
% TODO Visualization?!
\begin{description}
	\item[Completeness] Which programs of the test suite are covered by the solution.
	\item[Performance] How long does your solution need to optimize/transform the programs.
		How much memory does your solution need?
	\item[Conciseness] How many rules/lines/words/graphical elements do you need?
	\item[Purity] Is your solution entirely made of graph transformations? Do you need imperative/functional/logical glue code? What is the relationship between the graph transformation and the conventional programming part?
\end{description}

\nocite{*}
\bibliographystyle{eptcs}
\bibliography{case}
\end{document}